\documentclass[prd,article,showpacs]{revtex4}
\usepackage{amsmath}
\usepackage{graphicx}
\usepackage{epsfig}
\usepackage{dcolumn}
\usepackage{bm}
\usepackage{dcolumn}
\usepackage{slashed}

\providecommand{\U}[1]{\protect\rule{.1in}{.1in}}
\providecommand{\U}[1]{\protect\rule{.1in}{.1in}}
\providecommand{\U}[1]{\protect\rule{.1in}{.1in}}
\providecommand{\U}[1]{\protect\rule{.1in}{.1in}}
\providecommand{\U}[1]{\protect\rule{.1in}{.1in}}

\begin{document}

\title{Exploring effects of strong interactions
in enhancing masses of dynamical origin \bigskip}

\author{Alejandro Cabo Montes de Oca}

\affiliation{$^{*}$\textit{\ International Centre for Theoretical
Physics, \ Strada Costiera 11, \ Trieste, Italy }}

\affiliation{$^{**}$\textit{\ Grupo de F\'{i}sica Te\'{o}rica,
Instituto de Cibern\'{e}tica, Matem\'{a}tica y F\'{i}sica, Calle E,
No. 309, Vedado, La Habana, Cuba}\bigskip}

\begin{abstract}
\noindent A previous  study of the dynamical generation of masses in
massless QCD is considered from another viewpoint. The quark mass
is assumed to have a dynamical origin and is substituted by a scalar
field without self-interaction. The  potential for the new field
background is evaluated up to two loops. Expressing  the running
coupling in terms  of the scale parameter $\mu$, the potential
minimum is chosen to fix  $m_{top}=175$ $GeV$ when $\mu_0=498 $
$MeV$. The second derivative of the potential predicts a scalar
field mass of $126.76$ $GeV$. This number is close to the value
$114$ $GeV$ which,  preliminary data taken at CERN, suggested to be
associated with the Higgs particle. However, the simplifying
assumptions  limit the validity of the calculations done, as
indicated by the large value of $\alpha=\frac {g^2}{4\pi}=1.077 $
obtained. However, supporting statements about the possibility  of
improving the scheme come from the necessary inclusion of weak and
scalar field couplings and mass counterterms in the renormalization
procedure, in common with the seemingly needed consideration of the
massive W and Z fields, if the real conditions of the SM model are
intended to be approached.
\end{abstract}
 \maketitle

\section{Introduction}

The understanding of the hierarchy in the particle mass spectrum
remains to be a central question in High Energy Physics
\cite{nambu,fritzsch,coleman, bardeen,miransky,minkowski,clague}. \
In former works we had been considering an alternative perturbation
expansion for QCD, in which quark and gluon condensates are
incorporated in the free vacuum state for afterwards determine the
alternative modified Wick expansion \cite
{mpla,prd,epjc,epjc1,jhep,epjc2,hoyer,hoyer1,hoyer2,epjc19}. \ A
main motivation for this study was the wonder about what could
result to be the final strength of  dynamically generated quark
condensates in massless QCD, a theory in which the free vacuum is
highly degenerated, and the underlying forces are the strongest ones
in Nature \cite{coleman}. \ Another source of interest was to
imagine that the modified expansion could represent a foundation of
the similarities of the particle mass spectrum with the ones in
superconductor systems, as stressed in Ref. \cite{fritzsch,
fritzsch1}. This possibility was indicated, on one side, by the fact
that the employed fermion condensate free vacuums closely resembled
the similar ''squeezed'' states structure of the usual BCS
wavefunctions in standard superconductivity theory \cite{prd}.
  In \ Ref. \cite{mpla,prd,epjc,epjc1,jhep,epjc2,epjc09} some indications about
the possible dynamic generation of quark and gluon condensates had
been obtained. However, although the evaluated corrections signaled
the dynamical generation of the quark condensate parameter, the
vacuum energy turned out to be unbounded from below as a function of
this quantity  in the two loop calculation done in \cite{epjc09}. In
this same work we restricted the discussion to the simpler case in
which there is only quark condensates present in the system. In this
situation, the Green's functions generating functional $Z$ was
transformed to a more helpful form, as the same functional integral
associated to massless  QCD, in which all the effects of the quark
condensates are now embodied in only one additional  vertex having
two quark and two gluon legs. This representation allows to
systematize the diagrammatic expansion. In particular it permits to
implement the dimensional transmutation effect. The evaluation of
next corrections of the potential in order to search for its
stabilizing minimum are expected to be further considered. These
next corrections, however, show complications for their calculation
due to the taquionic character of the ladder approximation for the
gluon propagator employed  up to now in the evaluations.

\ In the present work we intend to also start examining the same
problem from another angle. The first motivation for doing this came
after taking into account that the fermion condensation properties
of massless QCD should be described by the effective action for composite
operators, in particular the $%
\overline{\Psi }(x)\Psi (x)$\ one \cite{effcomp1,effcomp2}. \ We
also plan to consider this approach in the near future. However, a
closely related, but simpler to tract task, can be to consider the
same generating functional of the $\overline{\Psi }(x)\Psi (x)$\
insertions for massless QCD
(which Legendre transform \ determine the mean $\left\langle \overline{\Psi }%
(x)\Psi (x)\right\rangle ),$ but in which the sources $J(x)$ in
place of being external and auxiliary ones, are chosen to be
dynamical quantities. \ This procedure then,  amounts to investigate
the same question about the dynamic generation of the quark masses,
but through the interaction of massless QCD with another dynamical
system. \ At this point, it can be observed that it is an
increasingly accepted viewpoint, that all the coupling constants and
mass parameters can have a dynamic origin \cite {witten,polchinski}.
Then, two possibilities for the physical relevance of introducing
the dynamical mass come to the mind. One is the interaction with an
scalar field in which the quark mass (source) is given through
Yukawa terms. \ The same SM is included in this case, but for the
restrictive condition of the scalar field system to show a negative
squared mass \cite{musolf, casas1,nielsen}. \ Another interesting
situation of a dynamical quark mass, comes from superstring theory,
where it is known that the fluxes in the compactified spaces
generalizing the Kaluza-Klein approach, can produce massive terms
for fermions in their low energy effective actions. This happens by
example, in Type IIB superstrings \cite {witten}. \  In this work we
decided to consider the first variant and assume, for the sole sake
of simplicity, that the quark interacts with a scalar field which
mass and potential terms in the classical Lagrangean  are both zero.
As it will be  seen, We plan to delete this calculational
restrictions, as a consequence of the obtained conclusions. \ The
initial exploration consists in evaluating the effective potential
of massless QCD as a function of the ''mass'' field background up to
the two loops approximation. This is decided with the aim of
answering  the question about whether or not at some special and
reasonable renormalization point conditions, the strong QCD
interactions becomes able to generate a minimum
of the potential determining the measured top quark mass of nearly $m_{top}=173$ \ $%
GeV.$

\ The results for the one loop potential indicate that this
contribution becomes unbounded from below at high ''mass'' field
values. The two loop terms, even after assumed to have  small
coefficients, dominate at high field values, determining a minimum
at some given mass value depending on the parameters. This minimum
is higher than $m_{top}$ if the strong coupling is chosen to be
small, that is, not in the infrared region. It is an interesting
fact, that this minimum seems to be related with the known existence
of a second minimum of the Higgs potential at large values of the
Higgs field. This high field minimum is precisely produced by the
contribution of the Yukawa term for the top quark, which has the
same form as the one considered here \cite{musolf,casas1,nielsen}.
This high field minimum appears in the evaluations associated to the
Standard Model (SM), at high energy renormalization point where the
strong interactions are small. However, as it will be remarked
below, the initial evaluations done here suggest an alternative
picture for implementing the symmetry breaking in  the SM model in
which a hidden  relevance of the strong coupling can exists. \

The work proceeds by evaluating the potential at the values of the
running coupling satisfying the one loop RG dependence on the scale
parameter $\mu $ by assuming the estimate  $\Lambda _{QCD}=217$
$MeV.$ \ In this circumstance, the minimum of the effective
potential as a function of the ''mass'' field can be fixed at
$m_{top}\approx 175$ $GeV,$ after the scale parameter is chosen at
the value $\mu _0=498$ $MeV.$ In addition, the second derivative of
the effective potential in its minimum at $\phi \sim 175$ $GeV$, \
gives  a result of $\ \ m_{\phi}$ $\sim 126.76$ $GeV$ for the scalar
field mass. It should be noted that the resulting value of the
coupling $\alpha =1.077 $ becomes neatly inside the non perturbative
region, and therefore the validity of the evaluated quantities in
this first exploration is not assured.
 However, after including the ''mass'' field in a two loop RG
 improvement of the evaluated effective action, it is imaginable that the potential can be
 made RG invariant up to higher $\mu $ scales,  where the strong coupling
 becomes low, but the scalar and weak ones are enhanced. Therefore, the results could
 become able to imply the  generation of the  top quark mass from a model in which
 the mass has a dynamical origin.
 This possibility seems feasible after taking in mind some considerations:
 1)  That a $\lambda \phi ^4$ counterterm should be forced to be included
 after a consistent renormalization procedure is developed. Therefore,
 further terms of similar analytical structure as functions of the
 "mass" field  are expected to appear in the improvement of
the simple analysis done here. In particular the coefficient of the
squared logarithm of this field \ terms (which makes the potential
bounded from below) should become a linear combination of $g_0^2$
and  $\lambda $.
 2)  Further, it can be noticed that our  aim is to expect a contact
 with the real SM model, where the scalar field is in fact a doublet
 and tightly interacts  with the also very massive Z and W bosons.
 Therefore, for the model to becomes realist, it seem necessary to
  also include  the contributions to the effective action  of all such
  additional modes.  The large masses of the Z and W particles
  should determine appreciable contributions to the potential, and
  the non asymptotically free dependence with the scale $\mu$ of the
  weak coupling can be expected to help in assuring the
  renormalization group invariance of the potential as described in the next point.

3) Thus, since the gauge field coupling  is asymptotically free, but
the weak and scalar ones  are  not, it can be expected  that under a
reduction of the scale $\mu $, the $g_0^2$ and the non
asymptotically $\lambda $ and weak couplings  can interchange their
roles in maintaining the RG invariance of the potential. Therefore,
assumed this effect to occurs up to a point in which the QCD
coupling is able to dominate, it can be possible to interpret that
the symmetry breaking effect in the SM model might be also
considered as strongly linked with the strong interactions dynamics
of QCD.  As remarked, the strong links that the RG could imposes on
both kind of fields interacting through the chirality breaking
Yukawa terms,  could be the main element determining the above
mentioned  hidden role the strong interactions.  Curious and perhaps
related  deep links between the gauge theories and Higgs model were
also identified before in Ref. \cite{nielsen}.

Therefore, in general terms,  the results suggest the possibility to
reformulate the SM model  by deleting the negative mass in the
classical potential of the Higgs field, and  fixing the dynamically
generated minimum of the scalar ''mass'' field to determine  the
value of the  top quark mass. \ The further investigation of this
question along the proposed lines is expected to be considered
elsewhere.

Finally, the above remarks also leads to conceive  that the quark
masses could be in essence, dynamically generated by the strong
forces, through the Yukawa terms (associated to their fermion fields
components) which are generated by the fluxes in the low energy
action of superstring theories. \  In this view the Higgs field
could be realized  by the flux created  mass factors of the top
quark chiral condensate operator in the low energy actions.

\ The work proceeds as follows.  In Section II the action and the
notations to be employed for the generating functional of massless
QCD in interaction with the "mass" scalar field will be defined. \
Section II will consider the MS evaluations of the one and too loop
contribution to the effective potential. In Section IV a discussion
about the above remarks, following from the examination of the
evaluated potential will be presented. The results are again
commented in the Summary.

\section{QCD interacting with a dynamical ''mass'' field}

Let us consider the Green's functions generating functional \ \ $Z$
describing massless QCD fields in interaction with a scalar field $\phi ,$
which  is assumed to be  real  for the sake of simplicity. The functional
has the form
\begin{equation}
Z[j,\eta ,\overline{\eta },\xi ,\overline{\xi },\rho ]=\frac 1{\mathcal{N}%
}\int \mathcal{D}[A,\overline{\Psi },\Psi ,\overline{c},c,\phi ]\exp
[i\text{ }S[A,\overline{\Psi },\Psi ,\overline{c},c,\phi]],
\label{Z}
\end{equation}
In this expression, $S$ is the action of the system closely following the
conventions of Ref. \cite{muta}:
\begin{align}
S& =\int dx(\mathcal{L}_0\mathcal{+L}_1\mathcal{)}, \\
\mathcal{L}_0& =\mathcal{L}^g+\mathcal{L}^{gh}+\mathcal{L}^q+\mathcal{L}%
^\phi , \\
\mathcal{L}^g& =-\frac 14(\partial _\mu A_\nu ^a-\partial _\nu A_\mu
^a)(\partial ^\mu A^{a,\nu }-\partial ^\nu A^{a,\mu })-\frac 1{2\alpha
}(\partial _\mu A^{\mu ,a})(\partial ^\nu A_\nu ^a), \\
\mathcal{L}^{gh}& =(\partial ^\mu \chi ^{*a})\partial _\mu \chi ^a,
\label{S} \\
\mathcal{L}^q& =\overline{\Psi }(i\gamma ^\mu \partial _\mu )\Psi , \\
\mathcal{L}^\phi & =\partial ^\mu \phi \partial _\mu \phi,  \\
\mathcal{L}_1& =-\frac g2f^{abc}(\partial _\mu A_\nu ^a-\partial _\nu A_\mu
^a)A^{b,\mu }A^{c,\nu }-g^2f^{abe}f^{cde}A_\mu ^aA_\nu ^bA^{c,\mu }A^{d,\nu
}-  \nonumber \\
& -gf^{abc}(\partial ^\mu \chi ^{*a})\chi ^bA_\mu ^c+g\overline{\Psi }%
T^a\gamma ^\mu \Psi A_\mu ^a+\overline{\Psi }\Psi \phi .
\end{align}
The classical action of the scalar field has been also chosen to be
self-interaction free and massless. As mentioned in the
introduction, this restrictions only constitute an  assumption in
attempting to simplify the evaluations. It will be  expected to be
relaxed  in improving the present study after implementing a
renormalization procedure by incorporating counterterms and
additional terms in the action, which can directly lead to masses
and scalar field couplings as required by the structure of the
divergences. It should be said in advance that this improvement will
result to be essential for verifying the preliminary conclusions of
this first examination of the problem. Note also that the
interaction of the scalar field with $\phi $ is simply a Yukawa
term, like the same one for the quarks in the Standard Model. \ \

\section{One and two loop effective potential of the ''mass'' field}

\ Let us consider in this section the evaluations of the contributions to
the effective potential $v(\phi )$ which QCD ''creates'' on the scalar field
$\phi ,$ up to the second loop order. The case of  homogenous values of $%
\phi $ will be assumed.

\subsection{ One loop term\ }

\ The analytic expression for the one loop contribution shown in Fig. 1 is
given by the classical logarithm of the fermion quark determinant as:
\begin{eqnarray}
\Gamma ^{(1)}[\phi ] &=&-V^{(D)}N\int \frac{dp^D}{i\text{ }(2\pi )^{D}}%
Log[Det\text{ }(G_{ii^{\prime }}^{(0)rr^{\prime }}(\phi ,p))], \\
D &=&4-2\epsilon,   \nonumber
\end{eqnarray}
\begin{figure}[h]
\includegraphics[width=5cm]{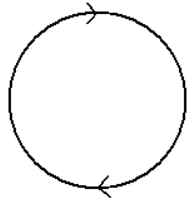}
\caption{ The figure shows the quark one loop correction. The result
depends on the scalar ''mass'' field $\phi $ through the quark free
propagator which is the usual free Green function of QCD, in which
the mass is substituted by $\phi $.} \label{fig1}
\end{figure}
where $D$ is the space dimension of dimensional regularization and the free
fermion propagator is written in the conventions of Ref. \ \cite{muta},
which will be used throughout this work. Its expression is given by
\begin{align}
G_{ii^{\prime }}^{(0)rr^{\prime }}(\phi ,p)& =\delta ^{ii^{\prime }}{\Large (%
}\frac 1{-p_\mu \gamma ^\mu +\phi }{\Large )}^{rr^{\prime }}  \nonumber \\
& =-\frac{\delta ^{ii^{\prime }}}{p^2-\phi ^2}{\Large (}p_\mu \gamma ^\mu
+\phi {\Large )}^{rr^{\prime }}.  \label{mqdef}
\end{align}
Assuming the case of QCD with $SU(N)$ symmetry for N=3, and
evaluating the spinor and color traces, the one loop expression can
be simplified to the form
\[
\Gamma ^{(1)}[\phi ]=V^{(D)}\frac N2\int \frac{dp^D}{i\text{ }(2\pi )^{D}}%
Log[(\phi ^2-p^2)^4].
\]
Taking the derivative over $\phi ^2$ of $\Gamma ^{(1)}[\phi ]$ allows to
write the easily integrable expression
\[
\frac d{d\text{ }\phi ^2}\Gamma ^{(1)}[\phi ]=V^{(D)}2N\int \frac{dp^D}{%
\text{ }(2\pi )^{D}}\frac 1{(p^2+\phi ^2)},
\]
which after making use of the formula
\begin{equation}
\int \frac{dp^D}{\text{ }(2\pi )^{D}}\frac 1{(p^2+\lambda ^2)}=\frac{%
\Gamma (1-\frac D2)}{(4\pi )^{\frac D2}}(\lambda ^2)^{\frac D2-1},
\label{oneloop}
\end{equation}
and integrating the result back over $\phi ^2$, results in the dimensionally
regularized expression
\begin{eqnarray}
\Gamma ^{(1)}[\phi ] &=&V^{(D)}\frac{2N\Gamma (1-\frac D2)}{(\frac D2)(4\pi
)^{\frac D2}}(\phi ^2)^{\frac D2-1}  \nonumber \\
&=&V^{(D)}\frac{2N\text{ }\Gamma (\epsilon -1)}{(\frac D2)(4\pi
)^{2-\epsilon }}(\phi ^2)^{2-\epsilon }.
\end{eqnarray}
Now, it is possible to divide $\Gamma ^{(1)}[\phi ]$ by
$\frac{V(D)}{\mu ^{2\epsilon }}$, in order to obtain the action
density. The quantity $\mu $ in the denominator is the dimensional
regularization scale parameter, and the divisor  $\mu ^{2\epsilon
},$ which tends to one on removing the regularization, is introduced
in order avoid results containing logarithms of quantities having
dimension. Then, the one loop Lagrangian density takes the form
\begin{equation}
\gamma ^{(1)}[\phi ]=\frac{\Gamma ^{(1)}[\phi ]}{\frac{V(D)}{\mu ^{2\epsilon
}}}=\frac{2N\text{ }\Gamma (\epsilon -1)}{(\frac D2)(4\pi )^{2-\epsilon }}%
\phi ^2(\frac \phi \mu )^{-2\epsilon }.
\end{equation}
Further, deleting the pole part of this formula according to the Minimal
Substraction rule, and taking the limit $\epsilon \rightarrow 0,$ gives the
finite part of the action density as
\begin{equation}
\left[ \gamma ^{(1)}[\phi ]\right] _{finite}=\frac{3\phi ^4}{8\pi ^2}%
(-3+2\gamma +2Log(\frac{\phi ^2}{2\pi \mu ^2})).
\end{equation}
Finally, the one loop potential energy density is given by the negative of
the above quantity
\begin{equation}
v^{(1)}[\phi ]=-\frac{3\phi ^4}{8\pi ^2}(-3+2\gamma +2Log(\frac{\phi ^2}{%
2\pi \mu ^2})).
\end{equation}
It can be noticed that this potential density is unbounded from below for
increasing values of the scalar field. This is the main property determining
the dynamical generation of the field $\phi $.

\subsection{ Quark-gluon two loop term}

\ The two loop quark-gluon term to be considered in this subsection is
illustrated in Fig. 2. \ Again, after evaluating the color and spinor traces
its analytic expression can be written as follows
\begin{equation}
\Gamma ^{(2)}[\phi ]=-V^{(D)}g^2(N^2-1)\int \frac{dp^Ddq^D}{i^2\text{ }(2\pi
)^{2D}}\frac{D\phi ^2-(D-2)p.(p+q)}{q^2(p^2-\phi ^2)((p+q)^2-\phi ^2)},
\end{equation}
where $g^2$ is the QCD coupling constant in the dimensional regularization
scheme, which introduces the scale parameter $\mu $ according to
\begin{equation}
g=g_0\text{ }\mu ^{2-\frac D2}=g_{0\text{ }}\mu ^\epsilon .
\end{equation}
\begin{figure}[h]
\includegraphics[width=5cm]{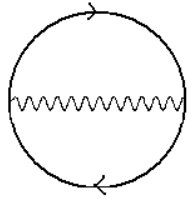}
\caption{ The two loop contribution determined by the strong interaction.
Again the $\phi $ dependence of the result is introduced though the free
quark propagator.}
\label{fig2}
\end{figure}
After \ symmetrizing the expression \ of  $\Gamma ^{(2)}[\phi ]$ under the
change of sign in the momentum $q$, by means of the integration variable
shift $\ p\rightarrow p-\frac q2,$ and the use of the identity
\begin{equation}
p^2=(p+\frac q2)^2-\phi ^2+\phi ^2-\frac{q^2}4-q.p,  \label{pcuadrado}
\end{equation}
the \ quark-gluon term may be written as the sum of two contributions of the
form \
\begin{equation}
\Gamma ^{(2)}[\phi ]=\Gamma ^{(2.1)}[\phi ]+\Gamma ^{(2,2)}[\phi ],
\end{equation}
in which $\Gamma ^{(2.1)}[\phi ]$ and $\Gamma ^{(2,2)}[\phi ]$  have the
formulae
\begin{eqnarray}
\Gamma ^{(2,1)}[\phi ] &=&-V^{(D)}2\phi ^2g^2(N^2-1)\int \frac{dk_1^Ddk_2^D}{%
i^2\text{ }(2\pi )^{2D}}\frac 1{k_1^2(k_2^2-\phi ^2)((k_1+k_2)^2-\phi ^2)}
\nonumber \\
&=&-V^{(D)}\frac{2\phi ^2g^2(N^2-1)}{i^2\text{ }(2\pi )^{2D}}J_{111}(0,\phi
,\phi ) \\
\Gamma ^{(2,2)}[\phi ] &=&-V^{(D)}\frac{(D-2)g^2(N^2-1)}{2i^2(2\pi )^{2D}}%
{\LARGE (}\int dk_1^D \frac 1{k_1^2-\phi ^2}{\LARGE )}^2,
\end{eqnarray}
where, the master two loop integral  $J_{111}(0,\phi ,\phi )$ , \ was
evaluated in  Ref. \cite{tausk}, and its explicit form for our particular
arguments is:
\begin{eqnarray}
J_{111}(0,\phi ,\phi ) &=&\int dk_1^Ddk_2^D\frac 1{k_1^2(k_2^2-\phi
^2)((k_1+k_2)^2-\phi ^2)}  \nonumber \\
&=&-\frac{A(\epsilon )\pi ^{4-2\epsilon }}{\epsilon ^2}(\phi
^2)^{1-2\epsilon }, \\
A(\epsilon ) &=&\frac{(\Gamma (1+\epsilon ))^2}{(1-\epsilon )(1-2\epsilon )}.
\end{eqnarray}
Then, it is possible to  write for $\Gamma ^{(2,1)}$
\begin{equation}
\Gamma ^{(2,1)}[\phi ]=-\frac{V^{(D)}2g_0^2\mu ^{2\epsilon }(N^2-1)}{\text{ }%
(2\pi )^{8-4\epsilon}}\frac{A(\epsilon )\pi ^{4-2\epsilon
}}{\epsilon ^2}\phi ^4(\phi ^2)^{-2\epsilon }.
\end{equation}
In the case of $\Gamma ^{(2,2)},$ the result is the square of the one loop
integral in equation (\ref{oneloop}), which leads to
\begin{equation}
\Gamma ^{(2,2)}[\phi ]=V^{(D)}\frac{g_0^2\mu ^{2\epsilon
}(N^2-1)2(1-\epsilon )}{\text{ }2(2\pi )^{8-4\epsilon }}\pi ^{4-2\epsilon
}(\Gamma (\epsilon -1))^2\phi ^4(\phi ^2)^{-2\epsilon }.
\end{equation}
Again dividing by $\frac{V(D)}{\mu ^{2\epsilon }}$ to evaluate the
action densities gives for the corresponding terms
\begin{eqnarray}
\gamma ^{(2,1)}[\phi ] &=&\frac{\Gamma ^{(2,1)}[\phi ]}{\frac{V(D)}{\mu
^{2\epsilon }}}=-v^{(2,1)}[\phi ]  \nonumber \\
&=&-\frac{2g_0^2(N^2-1)}{\text{ }(2\pi )^{2D}}\frac{A(\epsilon )\pi
^{4-2\epsilon }}{\epsilon ^2}\phi ^4(\frac{\phi ^2}{\mu ^2})^{-2\epsilon },
\end{eqnarray}
and
\begin{eqnarray}
\gamma ^{(2,2)}[\phi ] &=&\frac{\Gamma ^{(2,2)}[\phi ]}{\frac{V(D)}{\mu
^{2\epsilon }}}=-v^{(2,2)}[\phi ]  \nonumber \\
&=&V^{(D)}\frac{g_0^2\mu ^{2\epsilon }(N^2-1)2(1-\epsilon )}{\text{ }2(2\pi
)^{8-4\epsilon }}\pi ^{4-2\epsilon }(\Gamma (\epsilon -1))^2\phi ^4(\phi
^2)^{-2\epsilon },
\end{eqnarray}
which define the potential energy densities $v^{(2,1)}[\phi ]$ and $%
v^{(2,2)}[\phi ]$ as their negatives. \ After substracting the divergent
poles, taking the limit $\epsilon \rightarrow 0$ and numerically evaluating
all the constants in order to simplify the view of the resulting expression,
the total quark gluon two loop finite contribution to the energy density
reduces to the form
\begin{eqnarray}
v^{(2)}[\phi ] &=&5.34687\times 10^{-5}\phi ^4{\LARGE (}1227.16+474.276\text{
}g_0^2-  \nonumber \\
&&12.0(\text{ }29.6088+13.5253\text{ }g_0^2)\text{ }Log(\frac{\phi ^2}{\mu ^2%
})+  \nonumber \\
&&12.0\text{ }g_0^2\text{ }(Log(\frac{\phi ^2}{\mu ^2}))^2{\LARGE ).}
\end{eqnarray}
It can be seen that the leading logarithm term is positive,
indicating that the  contribution of the usual  QCD diagrams to the
potential up to the two loop approximation becomes bounded from
below as a function of $\phi $.

\subsection{Scalar-quark two loop term}

Finally, in this section let us present the two loop term being associated
to the quark-scalar loop illustrated in Fig. 3 . This calculation is similar
to the previous one, but simpler, due to the lack of spinor and color
structures in the vertices.
\begin{figure}[h]
\includegraphics[width=5cm]{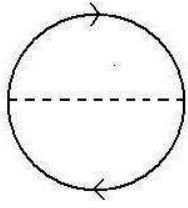}
\caption{ The diagram corresponding to the radiative corrections of the
''mass'' field $\phi $ in the two loop order. }
\label{fig3}
\end{figure}
This graph analytic expression can be written in  the following way
\ \
\begin{equation}
\Gamma _Y^{(2)}[\phi ]=V^{(D)}2N\int \frac{dp^Ddq^D}{i^2\text{ }(2\pi )^{2D}}%
\frac{p^2-\frac {q^2}{4}+\phi ^2}{q^2((p+\frac q2)^2-\phi
^2)((p-\frac q2)^2-\phi ^2)},
\end{equation}
which in a very close manner  as in the previous subsection, can be
represented as follows
\begin{eqnarray}
\Gamma _Y^{(2)}[\phi ] &=&V^{(D)}\frac{4N}{i^2\text{ }(2\pi )^{2D}}%
J_{111}(0,\phi ,\phi )  \nonumber \\
&&-V^{(D)}N{\LARGE (}\int \frac{dk_1^D}{i(2\pi )^D\text{ }}\frac
1{k_1^2-\phi ^2}{\LARGE )}^2  \nonumber \\
&=&\frac{V^{(D)}4N}{\text{ }(2\pi )^{8-4\epsilon }}\frac{A(\epsilon
)\pi
^{4-2\epsilon }}{\epsilon ^2}\phi ^4(\phi ^2)^{-2\epsilon }  \nonumber \\
&&-V^{(D)}\frac N{\text{ }(2\pi )^{8-4\epsilon }}\pi ^{4-2\epsilon }(\Gamma
(\epsilon -1))^2\phi ^4(\phi ^2)^{-2\epsilon }.
\end{eqnarray}
The division by the volume $\frac{V(D)}{\mu ^{2\epsilon }}$ allows
to write for the action density and its negative: the potential one,
the formula
\begin{eqnarray}
\gamma _Y^{(2)}[\phi ] &=&-v_Y^{(2)}[\phi ]  \nonumber \\
&=&\frac{4N}{\text{ }(2\pi )^{8-4\epsilon }}\frac{A(\epsilon )\pi
^{4-2\epsilon }}{\epsilon ^2}\phi ^4(\frac{\phi ^2}{\mu
^2})^{-2\epsilon }
\nonumber \\
&&-\frac N{\text{ }(2\pi )^{8-4\epsilon }}\pi ^{4-2\epsilon }(\Gamma
(\epsilon -1))^2\phi ^4(\frac{\phi ^2}{\mu ^2})^{-2\epsilon }.
\end{eqnarray}

The substraction of the divergent pole part in $\epsilon $, passing to the
limit $\epsilon \rightarrow 0$ and numerically evaluating the appearing
constants, gives for the potential density
\begin{eqnarray}
v_Y^{(2)}[\phi ]=-0.0000601522\text{ }m^4\left( 332.744-123.833Log[\frac{m^2%
}{\mu ^2}]+12.0 \, Log[\frac{m^2}{\mu ^2}]^2\right).
\end{eqnarray}
It can be noted that this contribution, being a two loop one, also
includes a squared logarithm term. However, its sign is contrary to
the one appearing in the quark gluon loop.

\section{ \ Discussion}

\begin{figure}[h]
\includegraphics[width=11cm]{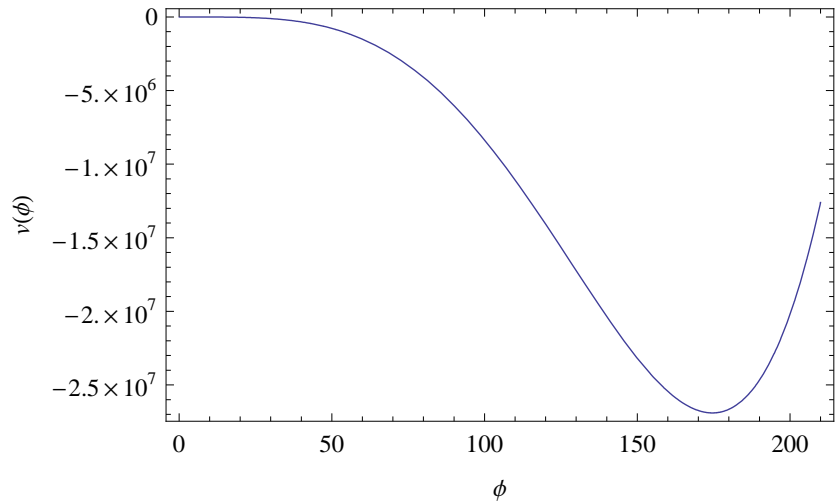}
\caption{ The effective potential for the ''mass'' field $\phi$ for
the value of the scale $\mu $ determining that the value of $\phi $
taken at the minimum potential,  is approximately equal to the top
quark mass $m_{top}~175\,\,GeV$. An interesting outcome is  that the
second derivative at the minimum is  $m_\phi =126.76\,\,GeV$, which
is close to  the energy $114$ $GeV$, at which  some  low statistic
data  measured at ALEPH experiment at CERN years ago,  led to
suspect its link with the Higgs mass. However, in the simplified
model in this work, the scale $\mu $ allowing the top mass fixation
is within the non perturbative infrared region: $\mu =0.498\,\,GeV$,
which gives a coupling value $\alpha =%
\frac{g_0^2}{4\pi }=1.077$. Thus, the described in the text
improvements of the treatment are needed for establishing the
possibility of fixing the observed  value of the top mass in the
scheme, and with it, the $ 126.76\,\,GeV$ estimation  of the Higgs
particle mass. } \label{fig4}
\end{figure}

Let us consider the sum of all the just evaluated contributions to
the potential energy density, defining it as the function $v(\phi
)$. Its expression is a combination of terms of the form $\phi ^4$,
$\phi ^4Log(\frac \phi \mu )$ and $\phi ^4(Log(\frac \phi \mu ))^2$,
with coefficients that only depend on the strong coupling $g_0$ in
the present first analysis. Then, in order to approach the physical
situation, we   evaluated the potential  $v(\phi )$ at the values of
$g_0$ satisfying the  one loop formula for the running coupling
constant \cite{muta}
\begin{equation}
g_0(\mu ,\Lambda _{QCD})=2\sqrt{\frac 27}\pi \sqrt{\frac 1{Log(\frac \mu
{\Lambda _{QCD}})}}.
\end{equation}
The  $\Lambda _{QCD}$ constant was chosen to be the estimate
$\Lambda _{QCD}=0.217\,\,GeV$. Next, we studied  the potential
curves  in order to examine the behavior of their minimum as
functions of $\phi $, when the scale $\mu $ is changed. It follows
that when the strong coupling starts to increase as the scale
diminish down to one $GeV$, the value of $\phi $ at the minima,
which determines the quark mass also decreases. For the particular
value of  $\mu =0.498\,\,GeV$ , the potential curve is shown in
Fig.\, \ref{fig4}.
The particular value of $\mu $ chosen,  fixes the position of the minimum at
a ''mass'' field $\phi $ defining a quark mass of $~175$ $GeV$. The set of
parameters for this curve are
\begin{eqnarray}
\mu  &=&0.498 \,\,GeV, \\
g_0 &=&3.684
  \,\,\,\,(\alpha =\frac{(g_0)^2}{4\pi }=1.077), \\
\Lambda _{QCD} &=&0.217\,\,GeV.
\end{eqnarray}
An interesting outcome is the fact that the squared root of the
second derivative of the potential curve in Fig. 1, estimates for
the mass of $\phi $ the value $m_\phi =126.76$ $GeV$. It can be
remarked  that this energy is close to the value of $114$ $GeV$ that
years ago appeared as signaling the presence of some low statistic
experimental data at CERN, and which were considered as perhaps
being related with the Higgs particle. This outcome, however, can
not yet be validated, as consequence of the resulting high value of
the strong coupling obtained. This elevated value might limit the
validity of the two loop and simplifying approximations employed
here. However, as remarked in the introduction, there exist
additional circumstances that indicate the possibility that the
picture remains valid after including renormalization and extending
the model in a consistent way.
 Let us recall  here a resume of these remarks:
 After implementing a renormalization procedure,  a $\lambda \phi ^4$ term
should necessarily be added to the Lagrangian (as a well as mass one
of the form $m^2\phi ^2$ ). This generalization of the here
considered simplified massless and non self-interacting form of the
scalar field Lagrangian, automatically should lead to additional
coefficients in the logarithmic expansion of the evaluated
potential.  Further, in order for  the extended theory to approach
the physical situation in nature, it seems necessary to also
incorporate, at least the also  very massive Z and W bosons and the
weak interactions among these fields. After that, in the same two
loop approximation considered here, the coefficient, let us call it
$f$,  of the
bounding from below squared logarithm term $(Log(\frac{%
\phi ^2}{\mu ^2}))^2$  can be expected have the form
\[
f=a_1\text{ }g_0^2+a_2\lambda+a_3\text{ }g_W^2.
\]
 That is, as a linear function  $g_0^2$, $\lambda $  and the squared
weak coupling $g_W^2$. Since the $a_1$ factor is positive as
evaluated here, \ in case that $a_2\lambda+a_3\text{ }g_W^2$ results
to be also positive, \ the minimum of the potential at the mass of
the top quark, has the chance of to be implemented  at a larger
scale $\mu.$ This possibility follows from the fact that, although
$g_0$ will diminish after increasing  $\mu, $ due to its
\emph{}asymptotically free character, the weak coupling $g_W$ and
the scalar field one $\lambda $ should be expected to grow because
they are  non asymptotically free. Therefore, this circumstance  can
be expected to  help in imposing the value of the observed value of
the top mass at a higher value of the  $\mu$ scale. If this
phenomenon occurs, it can be interpreted as a ''trading'' of the
roles of the weak interaction fields and the strong interaction
gauge fields upon passing from a high energy scale to the infrared
one. \ Thus, when working the perturbative theory at the infrared
scale, the Higgs mean field,  and with it the top, W and Z masses,
can then appear as generated by the  Yukawa interaction in
combination with the strong forces of the  quarks and gauge fields.
We expect to be able to support this picture in the extension of
this work.

At this point is of interest to observe that the appeared single
minimum seems to be related with the known existence of a second
minimum of the Higgs potential laying at large values of the Higgs
field. This minimum is recognized to be produced precisely by the
contribution of the top quark Yukawa term, which is of the same form
of the one considered here  \cite {musolf,casas1,nielsen}. \
Clearly, if the strong coupling is fixed at the perturbative scale,
the here obtained minimum will be shift to very high energies due to
the decreasing of the running coupling.

In conclusion, the remarks in this Section, support the possibility
that the inclusion of the usual ''$\lambda \phi ^4"$ coupling for
the ''mass'' field in the present discussion, but retaining a
positive value of its mass, in conjunction with the incorporation of
the weak interaction couplings and the  Z and W fields, could allow
the fixing of the value of top quark mass  at $175$ $GeV$, but  at
higher renormalization scale $\mu$. Note that after this
generalization, the structure of the leading logarithm terms
contributing  to the potential  will be similar, and moreover, the
unbounded one loop correction is not ruled out that could be able to
dominate the terms being  linear in the logarithm of $\phi$.
Therefore, the fixation of the position of the minimum of the
potential at the observed mass of the top quark, but at a  higher
scale $\mu$, could be eventually again approximately furnish a
reasonable estimate for the Higgs mass. The success in this task
could lead to a modification of the SM in which the so called second
Higgs minimum could substitute the classically imposed one in
defining the SM ground state. We expect to present elsewhere the
results of a current examination of these issues.

\section{Summary}

Previous investigations about the dynamical generation of large
quark masses in a modified version of PQCD were continued here but
starting from another viewpoint. For this purpose the mass parameter
of the quarks in QCD (or equivalently, a constant source in the
generating functional for the quark condensate composite operator in
massless QCD) is supposed to be a dynamical quantity. Examples of
such situation are the same Yukawa interaction in the SM model and
the masses generated by fluxes within compactified spaces in
superstring theories. \ A simple model of a real scalar field
without self-interaction is adopted in the initial exploration done
here. The effective action of massless QCD as a function of the
introduced ''mass'' field background, the strong coupling $g_0$ and
the scale parameter $\mu $, is calculated up to two loops by
employing the MS scheme. After imposing the one loop renormalization
group (RE) expression for $g_0$ as a function of $\mu $ and choosing
$\Lambda _{QCD}=217$ $MeV,$ the single minimum of the evaluated
effective potential can be fixed at a scalar field value
approximately equal to
the top quark mass $m_{top}\approx 175$ $GeV,$ for the scale parameter at $%
\mu _0=498$ $MeV.$ The second derivative of the effective potential
gives a scalar field mass of $126.76$  $GeV$. This value  is close
to the energy at which some insufficient experimental data taken at
ALEPH experiment at CERN years ago, suggested that might be
associated with the Higgs particle. However, in this first view of
the problem, the here employed two loop
expansion is yet inaccurate due to the large value of the coupling $\alpha =%
\frac{g_0^2}{4\pi }=1.077$. However, the identified possibility
motivates the search for validating the analysis. This task seems
promising by  developing a full renormalization scheme including
counterterms, the also very massive modes of the Z and W particles
and a RG improvement of the effective potential to be invariant up
to a higher renormalization scale. In this sense the following
possibility was here identified. Let us consider the introduction of
the usual $\lambda \phi ^4$ coupling for the ''mass'' field in the
present discussion. It should be noted that it will automatically
enter in the consistent renormalization scheme to be introduced.
Even a positive mass value of the renormalized Lagrangian mass of
$\phi $ should be taken into account. Moreover, let us also
introduce the contributions of the W and Z particles and their
associated weak couplings. Then, these modifications seem that could
be  able to allow the fixing of the top quark mass, in a similar way
as it was done here, but now at a higher energy scale $\mu $ at
which perturbation theory can be valid. This is suggested because
now there can appear further bounded from below squared logarithm
terms coming from the contributions of the Z,W and scalar four legs
vertices.
 Since the $%
g_0^2$ behavior is asymptotically free, but the weak and scalar
couplings are not, their influence  upon increasing $\mu $, could
result to be opposite. Then, it  might happens they compensate
between them, allowing in this way to fix the minimum of the
potential at the same $m_{top}=175$\, $GeV$ value, but now at a
higher $\mu$ scale.   The verification of this idea will support the
detection  of a Higgs scalar particle near the energy value
$m_{Higgs}=126.76$  $GeV$,  which the present exploratory study
identified.  We expect to consider these possibilities in the coming
extension of the work.

It was also underlined, that the discussion seems to be closely
related with the known existence of a second minimum of the Higgs
potential at large values of the Higgs field. This second minimum is
recognized to be created precisely by the contribution of the Yukawa
term for the top quark, which is of the same form as the one
considered here \cite{musolf,casas1, nielsen}.

In concluding, it can be also  remarked that the discussion also
motivates the speculative idea about that  the Higgs field could be
realized as a dynamical ''mass'' scalar field defined by the fluxes,
which naturally appears in the Yukawa terms of the low energy
effective action in superstring theories.

\begin{acknowledgments}
The author wish to deeply acknowledge the very helpful support and
funding  received from various institutions which allowed preparing
this work: The Caribbean Network on Quantum Mechanics, Particles and
Fields (Net-35) of the ICTP Office of External Activities (OEA), the
"Proyecto Nacional de Ciencias B\'{a}sicas" (PNCB) of CITMA, Cuba,
the High Energy Section of ICTP (Trieste, Italy), the Christopher
Reynolds Foundation (CRFNY, New York, U.S.A), the Departments of
Physics of Wisconsin and Cornell Universities (U.S.A.) and the
Institute of Physics of the Bonn University. I am also very much
grateful by the valuable comments and discussions with M.
Ramsey-Musolf, P. Fileviez, T. Han, A. Leclair, H. Tye, K. Dines, Y.
Dokshitzer, K. Tsumura,  A. Cabo-Bizet, N. G. Cabo-Bizet, G.
Thompson, H. P. Nilles, A. Klemm, G. von Gehlen, R. Flume, M. Drees,
M. Trapletti, etc.
\end{acknowledgments}

\end{document}